\documentclass[prb,preprint]{revtex4-1} 

\usepackage{amsmath} 

\usepackage{graphicx} 

\usepackage{multirow} 

\usepackage{array}

\usepackage{setspace}

\begin{document}

\title{Sixteen years of Collaborative Learning through Active Sense-making in Physics (CLASP) at UC Davis}
\author{Wendell Potter\cite{note1,note2}}
\author{David Webb\cite{note2}}
\author{Emily West\cite{note2}}
\author{Cassandra Paul\cite{note2,note3}}
\author{Mark Bowen\cite{note2,note4}}
\author{Brenda Weiss\cite{note2}}
\author{Lawrence Coleman\cite{note1}}
\author{Charles De Leone\cite{note1,note5}}
\affiliation{University of California at Davis, Department of Physics, Davis, CA 95616}
\email{webb@physics.ucdavis.edu} 
\date{\today}

\begin{abstract}
This paper describes our large reformed introductory physics course at UC Davis, which bioscience students have been taking since 1996. The central feature of this course is a focus on sense-making by the students during the five hours per week discussion/labs in which the students take part in activities emphasizing peer-peer discussions, argumentation, and presentations of ideas. The course differs in many fundamental ways from traditionally taught introductory physics courses. After discussing the unique features of CLASP and its implementation at UC Davis, various student outcome measures are presented showing increased performance by students who took the CLASP course compared to students who took a traditionally taught introductory physics course.  Measures we use include upper-division GPAs, MCAT scores, FCI gains, and MPEX-II scores.
\end{abstract}

\maketitle

\section{Introduction}
In 1995 two faculty members, a postdoctoral student, and several graduate students at the University of California, Davis (UC Davis) taught two trial versions of a new introductory physics course after experimenting for several years with radical modifications to the lab portion of the traditional course.\cite{PotterDeL}  The total replacement of the traditional course with the reformed course occurred during the 1996-97 academic year. Henceforth, all biological science majors at UC Davis have been taking the CLASP course to satisfy their requirement of a one-year physics course, unless previously satisfying this requirement prior to transferring to UC Davis.  The course developers were strongly influenced by research in science and physics education, particularly, the work of Rosalind Driver,\cite{DrivAsoko} Roger Osborne,\cite{OsbFrey} and Merlin Wittrock\cite{OsbWitt} in science education and by David Hestenes\cite{Hest87} in physics education, as well as by their extensive personal experiences in university teaching and collaborative work with K-12 teachers and K-12 educators/researchers. Their efforts were informed and strongly influenced by the constructivist idea that each student builds their own knowledge through discussion and argumentation. Toward this end these curriculum developers' specific goals were to develop a large-enrollment course that provided collaborative group work, that kept group discussions and argumentation at a high intellectual level,\cite{,Footnote,Bloom} and that kept the students' main focus on concepts rather than calculations, on genuine understanding\cite{Ausubel} and sense-making rather than rote memorization, and on the few big ideas in physics rather than the very many small details.  Essentially all aspects of the new course including the physics topics covered, the organization of those topics, the delivery of instruction, and the exam questions and the grading of those questions were significantly changed from the standard introductory physics course that the CLASP course supplanted.  In subsequent years many instructors made contributions to the course, which evolved as experience changed ideas of what could and should be accomplished with this group of students at this stage in their academic careers.\cite{footnoteA}   However, the general purpose and the general structure of the three-quarter, one-year course has been consistent throughout its existence. This paper describes the CLASP course, which is officially listed as Physics 7A, 7B, and 7C at UC Davis; we will typically refer to the course generically as our CLASP (Collaborative Learning through Active Sense-making in Physics) course/program.

At UC Davis, CLASP is a large course with $\sim 1700$ students enrolling in the course each academic year.  There are currently five independent course sections of each of the three parts of the course, CLASP A, CLASP B, and CLASP C, taught during each academic year with approximately 300 students in each course. Two faculty instructors and five graduate-student teaching assistants teach each separate course.  The CLASP course has essentially the same student-instructor ratio as found in the traditionally taught course for physics and engineering majors at UC Davis. CLASP shares a couple of features with another large enrollment course reform effort SCALE-UP,\cite{ScaleUp} that was independently being carried out at about the same time.  Both course development efforts were responses to the desire to incorporate extensive collaborative group work in large enrollment courses, similar to that being done in smaller-enrollment studio-style courses\cite{Studio}. In both CLASP and SCALE-UP lecture mode is minimized with students working primarily in small groups for four to six hours per week; however, in CLASP students are in sections of 25 to 30 students working at tables of five students, while in SCALE-UP students are in rooms with many more students, up to 90, working at tables of nine students. In this respect, CLASP is more similar to studio-style courses than to SCALE-UP. In addition to the fact that SCALE-UP was developed for the engineering/physics course and CLASP for the physics course taken by biological science majors, there are major differences in many other aspects of the course. These differences will become apparent in the following description of the course.

The paper begins with a discussion of the general structure and pedagogy of the course.  We then include sufficient detail of the actual course to illustrate the class atmosphere as experienced by an instructor or a student.  Finally, we include a summary of a few ways that we have used to assess the course .  In particular we discuss our students' learning of physics, their understanding of physics and science in general, and their performance in subsequent courses.  These measurements lead us to conclude that the CLASP courses are a significant improvement over the series of courses that were replaced.

\section{General Features of CLASP}
\subsection{Content organization}
\subsubsection{Organized around a set of models that address large classes of phenomena}
One of the most distinct differences between CLASP and the course it replaced, as well as with the vast majority of both reformed and traditional introductory physics courses, is in the order and organization of the physical ideas that we include in the curriculum. In traditional courses content is typically presented as a sequence of topics, each of which addresses a small, albeit important, element of a larger organizing structure. For example, a table of contents of a typical traditional introductory physics textbook might list 30 or more chapters with each chapter divided into five to ten or more topics. Students' focus is typically directed to each topic serially by the textbook, the lecture, and the assigned homework problems. The goal is to have students master the particular construct or concept or understanding of a particular phenomenon addressed within that topic before moving on to following chapters where these separate elements are logically combined into a more encompassing theoretical structure. For example, students first master the elements of kinematics, position, velocity, and acceleration, including their vector nature and relationship to each other, before encountering Newton's 2nd Law. After mastering the fundamentals of applying Newton's laws to a rather tightly constrained set of physical phenomena involving only constant acceleration, the logical structure is eventually extended to include the major conservation laws of energy and momentum.  This type of course may not be as useful as one might hope to students\cite{Kim02} who are learning the subject for the first time and who are judging ``what is important to really understand'' by the number of problems that they have to solve using the various ideas and by the overwhelming number of algorithms that seem important.  As noted above, the originators of CLASP wanted to keep each student's focus on the main ideas of physics (i.e. exactly on ``what is important to really understand'') and the unity inherent in the structure of physics rather than on the immense number of detailed specific examples and algorithms that are inevitably used to show how these main ideas play out in the world.\cite{Hest87}  Toward this end, the course is organized around a set of 27 models\cite{Mag02} that correspond to the sets of ideas physicists use to understand and make sense of many of the major features of the physical phenomena these students will encounter in their courses and careers.  Of these models, it is probably fair to say that about a half dozen of them are the most important overarching models.\cite{Footnote01}  These models, and this organization of ideas, are prominent in all of the work that the students do so we will briefly describe their location in the course. The term ``model'' as used in CLASP frequently does not refer to historically defined sets of ideas and relationships among those ideas that have been given the name model, but rather as the collection of ideas and the relationships among those ideas, which, when grouped together, prove useful to these students as they make sense of, develop explanations of, and make predictions of phenomena relevant to their needs.\cite{Brewe}

\subsubsection{Selection and ordering of models in CLASP}
In an attempt to build on the students' familiarity with chemistry, the series of courses begins with \emph{conservation of energy} (both internal energies and mechanical energies).  This is immediately followed by the \emph{statistical properties of systems of large numbers of atoms} and so completes the discussion of \emph{Thermodynamics}.  Next, conservation of energy ideas are used to analyze \emph{fluid flow} and \emph{electrical charge flow}.  After this is a shift to two other conservation laws, \emph{conservation of momentum} and \emph{conservation of angular momentum}, which also serves as the introduction to \emph{Newtonian mechanics}.  Following these discussions in mechanics, we introduce \emph{wave models}, \emph{interference}, and \emph{optics}.  Finally, the students discuss \emph{fields} (mainly E\&M) and \emph{quantum mechanics}.  The full list of models, in order, is shown in Table \ref{ModelList}.

\begin{spacing}{1.0}
\begin{table}
\caption{\label{ModelList}Models used in CLASP (1 academic year)}
\begin{tabular}{|>{\centering} m{150pt} |>{\centering} m{150pt} | m{150pt} <{\centering}|}
\hline Part 1 - 1st Quarter  &  Part 2 - 2nd Quarter  & Part 3 - 3rd Quarter \\ \hline \hline
Three-Phase Model of Matter & Steady-State Energy-Density Model (fluids, electric circuits) & Plane Wave Model \\ \hline
Energy-Interaction Model & Linear Transport Model (fluid \& charge transport, heat flow, diffusion)  & Wave Superposition and Interference Model  \\ \hline
Intro Spring-Mass Oscillator Model & Exponential Change Model & Ray and Wavefront Model  \\ \hline
Intro Particle Model of Matter &  Galilean Space-Time Model &  Electric Charge and Electric Force Model  \\ \hline
Particle Model of Bond Energy &  Force Model &  Electric Field and Potential Model \\ \hline
Particle Model of Thermal Energy &  Momentum Conservation Model  &  Magnetic Fields and Force Model  \\ \hline
Intro Model of Thermodynamics &  Angular Momentum Conservation Model &  Electromagnetic Wave Model  \\ \hline
Ideal Gas Model &  Newtonian Model &  Quantum Wave-Particle Duality Model  \\ \hline
Intro Statistical Model of Thermodynamics &  Simple Harmonic Motion Model &  Quantum States Model  \\ \hline
\end{tabular}
\end{table}
\end{spacing}

One significant advantage of this kind of grouping of phenomena and theoretical ideas is that particular concepts such as velocity, for example, are dealt with only at the complexity required for that model and the associated phenomena. Thus, the notion of speed and the square of speed is all that is required in making sense of phenomena addressed with models related to conservation of energy.  It is not until midway through the course that the vector nature of velocity is needed when students encounter conservation of momentum models.  Similarly, initially in the course, force can be thought of simply as a push or a pull exerted by one object on another is a particular direction.  Concepts such as these are further developed when necessary in later models. This is in stark contrast to the linear, logical development in traditionally organized introductory physics curricula.

\subsubsection{Representational tools used to keep discussions focussed on ideas rather than equations}
\begin{figure}[h]
\centering
\includegraphics[width=6.5 in]{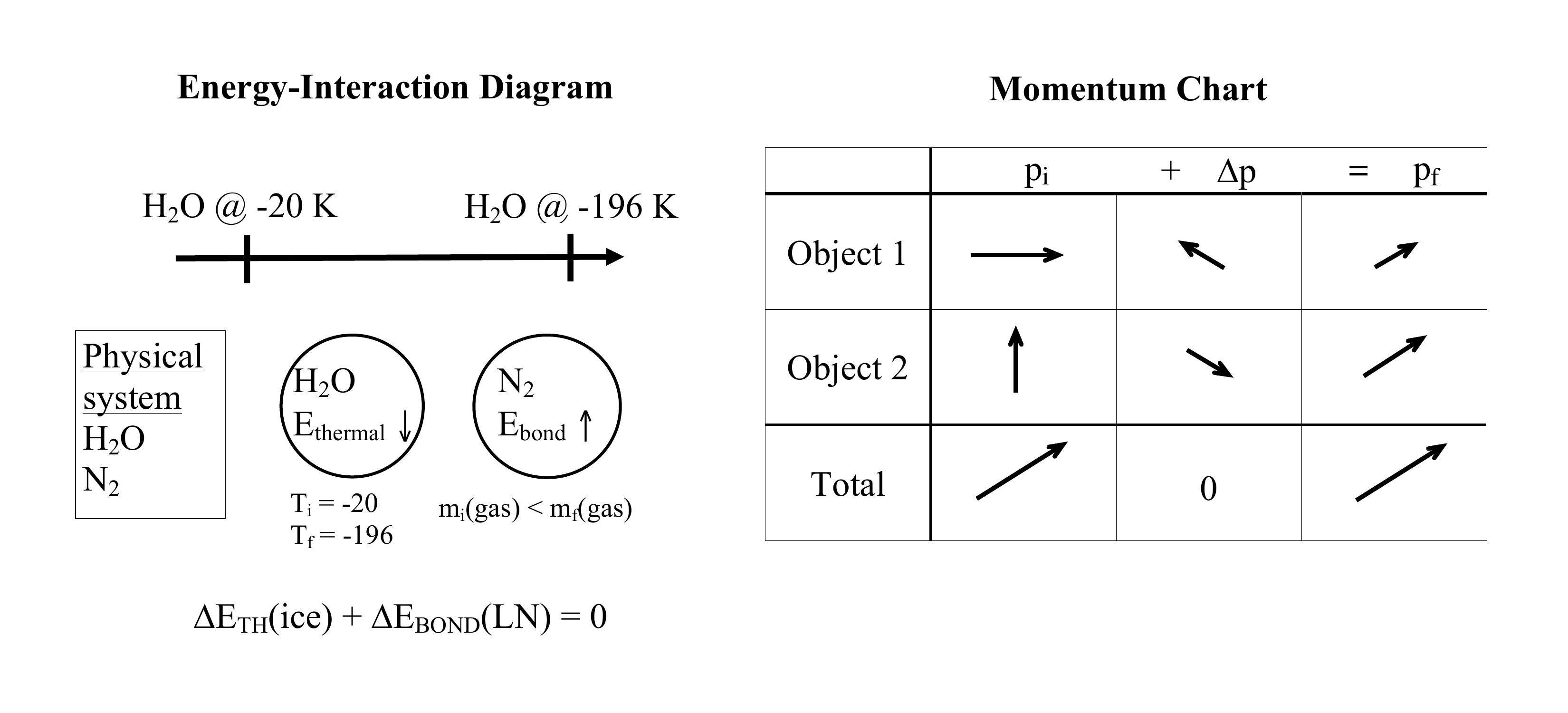}
\caption{\label{Pics}The diagram on the left is constructed by the students to help them think and talk about energy exchanges.  This particular example describes exchanges that occur when an ice-cube is added to a large container of liquid nitrogen.  The diagram on the right is constructed by the student to help them think and talk about the final motion of two objects (initially moving in different directions) that stick together after a collision (in this particular situation forces exerted on the two objects by their surroundings can be neglected compared to forces between the two objects during the interval of the collision).}
\end{figure}

One important reason for dealing so specifically with models in CLASP is the desire that the students who learn to work with these models end up building a conceptual structure that remains after they have completed the course that allows them to make at least some progress in understanding most physical situations they encounter in the real world. To help the students build and use this conceptual structure, most models rely heavily on diagrammatic or graphical representations that provide students a way to begin working with the model (i.e., begin understanding a particular physical situation) before\cite{Kohl07} writing down equations and doing complicated algebra.\cite{Wheeler}  The students readily use these diagrammatic and graphical representations in their discussions\cite{Stone12} and the representations clearly help them structure the presentations of their ideas to the entire class.  Two of these diagrammatic representations (one for energy conservation and one for momentum conservation) are shown in Figure \ref{Pics}.

\subsection{Course Structure}
A regular offering of a CLASP course at UC Davis includes one lecture section that meets once a week for 80 minutes and a discussion/laboratory (DL) that meets twice a week for 140 minutes each time. Weekly quizzes or biweekly exams are given in lecture, decreasing the actual time available for presentation. Thus, less than 1/4 of the in-class time is spent in lecture with the rest spent in the DLs, mostly in intellectually intensive discussions in small groups (typically 5 students), concerned with either i) making sense of the models or ii) using the models to make sense of various important physical situations.

\subsubsection{Lectures}
Lectures in the CLASP courses generally provide the first introduction of new material, but they do not have to be nearly as complete or as self contained as lectures in a standard physics.  For instance, in a CLASP lecture, the lecturer may define the appropriate technical words, describe the appropriate physical concepts and models, and ask the students to use them in real-world examples. However, the lecturer does not need to work out any example problems for the students, because the students will be working hard on applying the models in different (example) situations during their approximately 5 hours of DL time each week.  Many instructors consider the lecture time not taken up by quizzes or exams to be essentially a ``bonus time,'' rather than the time when the students must see all of the material. This is not only quite different from the usual view of lecture, it also liberates the instructor who is giving the lecture to engage the class in whatever activities the instructor thinks are most useful.  This is not a ``flipped'' classroom\cite{AshFlipped} because there is current no online component, but the lecture time can have some of the same features as lecture time in the flipped class.

\subsubsection{Discussion/laboratories}
The discussions in DL are what places our CLASP course in the category of ``interactive engagement" classes.\cite{Hake}  The classroom is arranged with students sitting around tables in groups of five with a wall-mounted black board near each table.  The group size was initially chosen for logistical reasons (fitting 25 students into lab rooms that had five bolted-down tables).  It soon became apparent that with this student population and the style and format of the activities the students engage in, five-member groups are optimal.  With a five-person group there is typically enough diversity of thinking among the five students to continually keep a discussion going.  Additionally, five students can comfortably engage each other while working at the table and when gathered around their black board.  The DLs currently meet in standard size lab rooms with six 4'x 6' blackboards around walls of the rooms, arranged so that there is a blackboard near each of the six small-group tables.  The tables are oval shaped with an equipment nub extending from the center of one of the long “sides.”  Lab-bench height tables with adjustable height stools seem preferable to lower tables with chairs, although we have used both and both work.  We prefer blackboards over whiteboards, simply because the blackboards hold up much better than white boards with the very heavy use they receive.  Other than the requirement to have a blackboard for each small group's use and a table that facilitates discussion among all five students, there are no other special constraints on the room as long as it can comfortably accommodate five or six small groups. We have found that if an instructor is responsible for more than six small groups, groups do not seem to get enough assistance from the instructor.  Classrooms with 8 or 10 groups and 2 instructors feel chaotic, and whole class discussions are not as productive as in the smaller classrooms. In a standard size lab room of approximately 800 to 1000 square feet floor space, five or six tables can be arranged comfortably and all groups can see each other's blackboards.  This is particularly important during the whole-class discussion periods discussed below.

The DL meetings are facilitated by an instructor who helps the student groups as they work independently and who also coordinates intermittent discussions involving the entire class.  The intent is that the pace of the DL is primarily controlled by the students, and that the discussions are carried out primarily in the students' voice, even when an instructor is present.  Students are provided with detailed prompts which guides their work.  During the initial four or five years, the student prompts were provided on a set of slides using an overhead projector.  After the curriculum became more stable, the prompts were printed in a workbook-style format.  We provide DL instructors with detailed instructor notes for each DL activity with frequent reminders to the instructor to be a ``guide on the side'' not a ``sage on the stage.''\cite{Footnote02}  However, approximately 30 different instructors are teaching discussion/labs in CLASP courses each year and there is large variation\cite{RIOT} among the instructors in how they interact with the student groups.  In spite of this variation, however, student performance as measured on quizzes and the final exam is rather independent of instructor interaction type.  What the students are actually doing during the DL is dependent to a much greater extent on the curricula materials provided to the students than on the actions of the individual instructor.

In a typical DL meeting the students will work through three activity cycles lasting 30 to 60 minutes each. Between the twice-weekly DL meetings students are assigned homework that builds directly on activities they have just carried out in the DL. Their homework is checked off at the beginning of the following DL meeting, not for correctness, but for effort at addressing the various prompts. Follow-up of the homework is often incorporated into the first or second activities in the next DL meeting. Thus, the homework is not an independent set of problems to get an answer to, but is integrally incorporated into the DL activities. A typical DL activity cycle typically consists of small group work, perhaps interrupted with one or two short whole class discussions, and a closing whole class discussion. These components are described in more detail below.

\paragraph{Small Group Work}
In the discussion/laboratory (DL) the students work in small groups on activities aimed at helping each student build their own personal understanding of the constructs of any particular model and the way in which these constructs are used within the model.  The activities are intended to help our students become fully literate in a particular model so the activities generally ask the students to discuss specific physical situations in their own words, discuss them using the technical words and concepts of the appropriate model, diagram the situation using one of the representations associated with the model, and, sometimes, translate these discussions into the mathematical language of the model.  Usually, once the translation into mathematical language is complete, solving the resulting algebra problem, substituting in the numbers, and doing the arithmetic is left for students to do outside of class.

As the students work, the instructor circulates the room listening to the conversations of each group and reviewing the content on the boards.  When the instructor sees or hears evidence that the students are stuck (perhaps two students are arguing at odds and cannot choose a way forward, or the whole group is heading down an unfruitful path), the instructor joins the group to help mediate a solution.  The instructor might ask the students to explain their application of the model to the situation at hand and remind the students of overlooked assumptions, or might use Socratic questioning to guide the student group back on track.  The specific instructor approach will vary depending on the particular problem the group is experiencing as well as instructors' beliefs about teaching.\cite{CassDiss, Speer, GoertzenBuyIn}

\paragraph{Whole Class Discussion}
The pace of the small-group discussions is determined approximately by the majority of the students in the class.  After a reasonable number of the small groups (half of them or more) have come to their conclusions about the activities that we asked them to work on, the instructor stops the small group discussions and leads a whole class discussion on the activity. Ideally, this whole class discussion is also carried out in the voice of the students (i.e., student-student discussions of the ideas).  A typical whole class discussion begins with one student presenting the group's work to the class.  The instructor might additionally or alternatively ask short summarizing questions of the class as a whole, ask students to review the content on all of the boards to discuss competing assumptions made by various groups, or prompt students to consider how the specific activity fits into the broader course goals.  Sometimes the instructors are provided specific prompts to use in the whole class discussions, but more typically the instructor uses his/her judgment in guiding the discussion.

We have many goals in our introduction of a discussion with the whole class at the end of an activity.  The first goal will be clear to any teacher.  The whole class discussion aims to leave each student, at a minimum, with a basic understanding of what ideas needed to be used in the activity, how they needed to be used, where these ideas fit into the field of Physics, and how the activities relate to other activities that they have done.  However, beyond this learning of physics concepts, we hope that our class gives our students a (somewhat) realistic view\cite{DrivAsoko} of how science proceeds and we see no reason that this cannot be done in concert with the first goal.  For instance, a whole class discussion may result in some groups advocating for one way of thinking about things and other groups advocating for another way (this is actually not uncommon when 5 or 6 groups work on their activities relatively independently) and then the discussion can bring out differing assumptions, differing viewpoints, and (of course) genuine conceptual misunderstandings.  The whole class discussion also gives the students a chance to practice developing their abilities to participate in proper scientific discussion and argumentation.\cite{Footnote03}  Engaging in and practicing authentic scientific practices such as argumentation helps students gain confidence in their ability to perform well on the quizzes/exams both in CLASP and in other science courses they subsequently take when faced with the task of applying the science (ideas, principles, models) to new phenomena. This is discussed in Section IV.

\subsubsection{Assessments of physics understanding}
We will discuss the details of assessments and grading in a separate paper but, briefly, the culture in the CLASP series at UC Davis is that there are many short quizzes, which are given in lecture. Most typical is a 20 - 25 minute quiz every week or a 30 - 35 minute quiz every two weeks. There is also a comprehensive two-hour final exam. There are two main reasons for very frequent quizzes.  The first is that the course emphasizes understanding of physical ideas and their application, so we want to give the students a way to monitor their understanding of each idea or set of ideas.\cite{BlackDylan}  The second is so that students have many chances to learn how to produce a scientifically correct argument and a complete discussion of a problem.

Finally, there is also a culture regarding the types of exam questions in the CLASP series at UC Davis that is followed by the majority of the instructors in charge of the 15 separate courses. This culture: i) values exam questions and problems that are significantly different from those that the students have already seen in the sense that they must apply the same model(s) to a completely new phenomenon; ii) values exam items that are not amenable to algorithmic solution; and iii) prizes the quality of a written scientific discussion (typically an explanation or prediction) given by a student above the algebraic correctness of a mathematical answer, although at times, both are required.

\subsubsection{Implementation Details}
At UC Davis over 1700 students complete the CLASP A-C series each year during the academic and summer sessions.  Most of these students are in bioscience/agriculture majors for which this physics series (or its equivalent) is a requirement of their major. There are five separate CLASP courses offered each quarter with 9 to 11 DL sections each for a total of about 50 DL sections meeting each quarter; each DL section meets twice each week, so approximately 100 140-minute DL meetings take place each week to accommodate the ~1500 enrolled students.  Each DL section has 25 to 30 students who nominally work in six groups of five students each. Each of these five courses is taught by two co-instructors along with four or five graduate teaching assistants (TAs); thus the entire CLASP program has 10 instructors and 20 to 25 TAs associated with it each term.  Usually 20 to 50\% of the instructors are regular faculty and the rest are either temporary lecturers or advanced graduate students who are known to be excellent CLASP TAs and who would like to gain broader teaching experience.

\paragraph{Co-Instructors}
The two co-instructors divide up the teaching times and responsibilities in any way they decide.  There are two identical lectures because the lecture room can hold only about half of the 300+ students. The most common way is for one instructor to give the two 80-minute lectures each week and to handle the major administrative duties of the class and for the other instructor to teach each of the two lead-off discussion/lab sections each week, run the two 50 min TA meetings following each lead-off DL, and deal with the administrative issues associated with the discussion/lab. Because no instructor teaches alone, it turns out that this course is a good way to introduce new instructors to teaching CLASP.  As discussed by Henderson et al.,\cite{HendersonCoTeach} this is also a natural way to promote instructional change because the course materials provide scaffolding that allows new instructors as well as more senior faculty to practice teaching an interactive class.  Co-teaching a CLASP course also provides a vehicle for an advanced graduate student to practice lecturing under the mentorship of an experienced faculty member teaching the course with him/her.  Both instructors are responsible for the final grades, so both work on writing and grading the quizzes/exams.

\paragraph{DL Instructors and TA professional development}
The role of the DL instructor (either a faculty member or a graduate student) is largely to facilitate the discussions that students have about their assigned activities and/or homework problems and to keep the students on task.  The actual DL activities are packaged with course notes and purchased by the students at the beginning of the quarter, but, sometimes, modified activities are distributed to the students during the quarter.

The DLs are where the students do much of their thinking and get much of their practice with the material, so they are the most important part of the course.  Graduate student teaching assistants (TAs) lead over 90\% of these DL sections and we have found that new graduate TAs must rapidly learn about both teaching and learning.  For this reason, we have a significant professional development program focused on our new graduate students who are teaching this course for their first time.  For new graduate TAs, this includes:

i) a mandatory 3-day introduction to teaching in CLASP, which, besides dealing with nuts and bolts of the job, also puts the new TAs in the roles of students working on CLASP activities, followed by putting the TAs in the roles of getting ready to teach DL, followed by putting the new TAs in the role of teachers (with reflections/comments on teaching and learning after each TA is finished).  These activities are interspersed with more general discussions of teaching and learning, which are taught in the CLASP (group discussion) style.

ii) a mandatory 1-hour per week TA training course during the first term that the new graduate student is enrolled at Davis and begins teaching a DL. This class is generally aimed at the theory and practice of teaching an interactive engagement type of class.\cite{Ishikawa}  Among other things, in this class the new TAs visit DLs of senior TAs and comment on what they have seen, discuss and practice grading, discuss teaching and learning, discuss the use of models in science, work on improving their teaching approach to whole class discussions as well as small group discussions, and monitor one meeting of one of their fellow new graduate student's class (using a computer program\cite{RIOT} to quantify how their fellow TA spends their time in class).

iii) attending the twice weekly one-hour TA meetings where both nuts and bolts issues related to the DL activities are discussed as well as more general issues related to teaching this type of course.

iii) We also offer (non-mandatory) TA professional development classes after the Fall term.  These are generally aimed at studying and improving each TAs teaching skills and/or the CLASP activities.

Goertzen notes that professional development activities may not be enough to affect the 'buy-in' of certain instructors, and that overall departmental norms may have a strong influence on how reformed courses are implemented by graduate student teaching assistants.\cite{GoertzenBuyIn}  The following section provides more information on how the the CLASP curriculum is viewed by the department.

\section{Outside Course Reviews and Physics Faculty Response to CLASP}

\subsection{\label{reviews}Objective Outside Reviews}
In 2006 the first two quarters of CLASP (Physics 7A and 7B) were separately identified as two of the top five examples of best practices in a national study of introductory college physics courses conducted by the Center for Educational Policy Research (CEPR) on behalf of the College Board. In a letter dated 9/1/2006 addressed to Winston Ko, UC Davis Dean of Mathematical and Physical Sciences, Professor David T. Conley, Director, CEPR, wrote: ``A total of 139 courses from across the nation were reviewed. In addition to being identified as being best practices overall, Dr. Potter's course was designated an, `exemplary practice' course by a panel of national experts in Physics. This is the highest designation awarded, and only a very few courses in our study met this standard of distinction and excellence. The study sought to identify best practices college courses that could inform the redesign of AP courses in Physics.''

\subsection{Sample Comments from UCD Faculty}
One feature of our CLASP course is that it introduces Physics faculty to interactive engagement classes.  This exposure could potentially affect the practice of instructors in other courses. Unfortunately, we don't have a measure of how much our faculty has been changed by the CLASP courses.\cite{Fairweather2010,HendersonCoTeach}  However, because the curriculum is pre-determined and the institutional constraints already handled by the developers, faculty who teach CLASP do not face many of the barriers that typically\cite{HendersonDancy} hinder the implementation of interactive engagement elements.  For example, Henderson and Dancy\cite{HendersonDancy} find that some faculty feel that the \emph{expectations of content coverage} are too great for a simple transition to interactive curricula.  Some also feel that their \emph{class size} and room \emph{layout} are not conducive to interactive environments, and still more do not like that they are challenging \emph{departmental norms}. The \emph{lack of instructor time} is also quoted as an issue. These particular barriers do not exist, or exist to a lesser extent for those teaching the CLASP curriculum. Thus it has been our experience that faculty who decide to teach CLASP courses have an overall positive experience.

In 2002, five faculty members who had taught the CLASP curriculum, but who had not been involved in its development participated in semi-structured interviews\cite{Weiss} about their experience in CLASP. The purpose of these interviews was not to evaluate CLASP, but instead investigate how instructors transitioned to teaching interactive courses.  The faculty were selected to participate in these interviews solely because they had taught the course during the academic year when the interviews were conducted.  It should however be noted that no faculty are forced to teach CLASP, if they do not wish to, and so all faculty in this sample chose to teach CLASP for one reason or another.

The faculty comments we present here is meant to provide insight regarding the transition experience of faculty teaching the CLASP course fairly early in its development.  It is not meant to be an evaluation of the CLASP course itself, but merely to indicate that faculty enjoyed implementing the interactive techniques. Note that in this dialogue, `Physics 7' refers to the CLASP course, and `Physics 9' refers to the traditional course.

In answer to the prompt: `What were your expectations coming into Physics 7' [\textit{CLASP}]?

Faculty member1 : \textit{I came into Physics 7 with a very negative attitude.  I \ldots had taught Physics 5} (the traditional intro-physics for bioscience course at UCD) \textit{three times through.  I had a student \ldots who would come in every day and tell me what a fiasco this whole Physics 7 was\ldots.  Rather than going to a faculty meeting and in complete ignorance try to stop this disaster, I figured the only honest thing to do was to try to teach it myself so that I could then draw my own conclusions.  When I did that I had sort of the complete opposite experience than} [my student] \textit{had.  I really enjoyed it.  I really felt that it was a much more dynamic learning environment\ldots.  Getting the students up presenting their responses to the class really forced them to think their ideas out very clearly.}

In answer to the prompt: `What did you find most surprising about teaching Physics 7' (CLASP)?

Faculty member 2:  \textit{How much fun it was.  It's a riot!  I mean you really get to meet \ldots six hours \ldots that's more time than most people spend with their kids at this age, you know, or younger.  And you get to know them all.  And that's kind of fun\ldots   it's the methodology of the activities.  How you work in little groups, and how the groups present their stuff to the larger group.  It's not the activity per se, but how we go about investigating it and sharing it (I hate that word) \ldots telling other people it.}

In a discussion following the prompt `What do you think of the attitudes that students have in Physics 7' (CLASP)?

Faculty member 3: \textit{The attitude} [of the students] \textit{may be unchanged, but because it forces them to talk and participate \ldots at least it draws out something in them that they don't get drawn out in a lecture class.  I was actually pleasantly surprised at what a large fraction would actually talk, would ask questions, puzzle on things\ldots.  I think that's why this is a successful class, because it forces some mental activity on the students’ part that is always lacking if they are sitting taking notes in lecture.}

On the subject of team teaching:

Faculty member 3: \textit{For a new instructor it was great, because it was so much less work for me, because all of you had experience and had figured out the subject matter and stuff like that.  I was thrilled, and I was happy to put in my two cents on the quizzes and things like that \ldots if I had had to write all the quizzes and the final, and all, it would have been an enormous amount of work for me.}

Overall, the faculty were very happy with the interactive engagement aspects of the course and all of them (except one who has retired) continue to teach it.  The main negative comments were regarding issues that have since been rectified (such as the lack of reference book), or had to do with the act of course reform in general.  Some of these comments are presented below for completeness.

Faculty member 3: \textit{I was expecting that the DLs would have been more finalized by the time it was handed to me two days before ldots.  That was a shock, that they weren't ready and they were in flux.}

Faculty member 2: \textit{Well, it's gotta' be polished.  It's been kind of a work in progress the whole time, it's now time to tighten it up a bit.}

Faculty member1: \textit{ \ldots polishing really means taking what you have and making incremental changes and there has been far too much wholesale throwing everything out.}

\section{Measurements of student learning and transfer}
In this section we will discuss some data that we have examined over the years and that help us judge some of the results of this course.  First, we directly compare students who took the CLASP series with those who took our previous physics series (Physics 5).  Then we discuss scores on concept inventories.  Finally, we discuss our students' general attitudes about physics.

\subsection{Direct comparison between CLASP students and Physics 5 students}
We hope the CLASP series better prepares students for later work.  In examining this possibility we use two different measures, students' work in later courses and students' MCAT scores.

\subsubsection{Preparation for later courses}
In the few years after the introduction of this CLASP course we had a chance to compare the students taking the CLASP series with those who took the previous UC Davis intro-physics series for bioscience students (Physics 5).  As a proxy for the upper division major GPA we calculate a student's GPA for the 7 quarters (just over two years) that preceded their graduation and use those to compare different groups of students.  We will call this GPA the UDGPA.  We only include students who had at least 65 quarter units (about 1.5 years of a normal class load) and we did not include any students who started their intro-physics series less than 5 quarters before their graduation.  Finally, we remove any intro-physics grade points and units that they received in the 7 quarters before graduation.

Bioscience students graduating in the years 1998 and 1999 had taken either the CLASP series or Physics 5 so, over those years, it's possible to compare the UDGPA across the two populations.  The results of these calculations are given in Table \ref{9899GPAs} and show that the students who took the CLASP series had higher UDGPAs.  A t-test shows that these Physics 5 and CLASP distributions are statistically significantly different, $p = 0.05$.  We also see that both males and females had higher UDGPAs at graduation if they took the CLASP series though t-tests show that neither of these was statistically significant, $p = 0.1$ for this test for females and $p = 0.55$ for males.  In this direct comparison we conclude that students who took the CLASP series performed better in their major courses than students who took the Physics 5 series.  One confounding aspect of this direct comparison is that the developers of the CLASP curriculum had tried ``interactive engagement'' activities in the laboratories of much of the Physics 5 series during 1993-95 so we estimate that at least 15\%-20\% of the students graduating in in this period\cite{Footnote05} had ``interactive engagement'' laboratory experiences.  Beyond this confounding issue, the direct comparison may be criticized on the grounds of a selection bias because these students have made a decision (either directly or indirectly) as to which Physics series to take.  For both of these reasons we do a second, somewhat different, comparison.

\begin{table}
\caption{\label{9899GPAs}Upper division GPAs for bioscience students graduating in years 1998 and 1999 who took physics at UC Davis.  Any intro-physics grades have been removed and quoted errors are standard error of the mean.}
\begin{tabular}{|>{\centering} m{60pt} |>{\centering} m{50pt} |>{\centering} m{70pt} |>{\centering} m{50pt} | m{70pt} <{\centering} |}
\hline Graduates in 1998 and 1999  &   Number who took Physics 5  & UDGPA of Physics 5 Students &  Number who took CLASP  & UDGPA of CLASP Students   \\ \hline \hline
All students & 779 & $3.093 \pm 0.017$ & $651$  & $3.142 \pm 0.018$ \\ \hline
Males  & 333  & $3.052 \pm 0.026$ & $217$ & $3.077 \pm 0.033$ \\ \hline
Females  & 446 & $3.123 \pm 0.022$ & $434$ & $3.174 \pm 0.022$  \\ \hline
\end{tabular}
\end{table}

As a check against the possibility of selection bias controlling the data discussed above, and also to get a cleaner measurement of the effects of CLASP, we compare students who graduated in 1993 and 1994 with those who graduated in 2000 and 2001.  There cannot be a selection bias here because students graduating in 1993 and 1994 could only have taken the Physics 5 series and, in addition, almost none of these students had any ``active learning'' laboratories in their Physics 5 classes.  Conversely, students graduating in 2000 and 2001 could only have taken the CLASP series.  Rather than compare these groups directly, we compare each of these groups of students who took UC Davis intro-physics to those students, graduating in the same majors, who did not take either Physics 5 or CLASP and so must have taken another intro-physics course\cite{Footnote50}. The vast majority of students in this second group have transferred into a bioscience major after two years at a community college; this is a sizable group of students at UC Davis (approaching 50\% of the upper-division students in many majors).  Since we are comparing ``4-year students'' in the biosciences to transfer students in the biosciences we will also compare 4-year students in non-bioscience majors to transfer students in the non-bioscience majors graduating in those same years so that we can decide if UC Davis transfer students became generally stronger or weaker between 1994 and 2001.  We calculate the UDGPA's for four groups: i) bioscience majors who entered as Freshmen\cite{Footnote05A}, ii) non-bioscience majors who entered as Freshmen, iii) bioscience majors who transfered from another college\cite{Footnote05B}, and iv) nonbioscience majors who transferred from another college.  The results are shown in Table \ref{FrshVSTransf} and one sees that transfer students had slightly higher (but not statistically significant) average UDGPA's than students from the same majors entering as Freshmen in both sets of years ($p=0.41$ and $p=0.65$) for non-bioscience majors and in 1993 and 1994 ($p=0.09$) also for bioscience majors.  We use the data for the non-bioscience majors as evidence that the strength of our transfer students did not change in those years.  It is notable that, of these four comparisons between four year students and transfer student, only bioscience majors graduating in 2000 and 2001 (those who took the CLASP series of courses) had higher average UDGPA's than transfer students in their majors and only for these two groups were the differences statistically significant ($p=0.0001$).  The magnitude of the UDGPA difference here is similar to that found in the direct comparison of Physics 5 and CLASP students graduating in the years 1998 and 1999 so we would argue that selection bias is not likely to have had a dominating effect on those data.

\begin{table}
\caption{\label{FrshVSTransf}Upper division GPAs comparing students entering as Freshmen with transfer students for two sets of graduation years and for two types of major (biosci or nonbiosci).  The quoted $p$-values are for the various t-tests comparing average UDGPA for 4-year students to that for transfers.  The non-bioscience majors did not have to take any physics courses and are included as a check against the possibility that the quality of our transfer students changed between those two sets of years.  Only students who took CLASP courses differed significantly from their transfer counterparts.}
\begin{tabular}{|>{\centering} m{70pt} |>{\centering} m{80pt} |>{\centering} m{55pt} |>{\centering} m{80pt} | m{70pt} <{\centering}|}
\hline Grad. Years  &  Type of Grad.  &  Number of students  &  UDGPA (Std. Error)  &  $p$-value from $t$-test  \\ \hline \hline
1993 \& 1994  &  NonBio Fresh.  & 1557  & 3.029 (0.012)  &  0.41 \\ \cline{2-4}
 &  NonBio Trans.  & 1378  & 3.043 (0.012)  &  \\ \cline{2-5}
 &  Bio Fresh.  & 566  & 3.039 (0.020)  &  0.65 \\ \cline{2-4}
 &  Bio Transf.  & 333  & 3.054 (0.026)  &  \\ \hline
2000 \& 2001  &  NonBio Fresh.  & 2037  & 3.028 (0.011)  &  0.09 \\ \cline{2-4}
 &  NonBio Trans.  & 1304  & 3.057 (0.013)  &  \\ \cline{2-5}
 &  Bio Fresh.  & 710  & 3.098 (0.017)  &  0.0001 \\ \cline{2-4}
 &  Bio Transf.  & 491  & 2.995 (0.021)  &  \\ \hline
\end{tabular}
\end{table}

Another example of increased student performance by CLASP students in subsequent courses was seen in a recent study supported by a NSF CCLI grant DUE-0633317 Improving the Learning Experience in Introductory STEM courses in a Large Research University. An unforeseen outcome of this study was significantly increased performance in the general chemistry course of a group of students who took the first quarter of the CLASP course during the first quarter of their freshman year before beginning the general chemistry course in the winter quarter. A control group started chemistry in their first quarter and CLASP in their third quarter. The performance of the control group in all three quarters of chemistry was marginally above the average performance of all students in the chemistry course ($\approx 380$), but not statistically significant ($p>0.05$). The performance of the students who took the first quarter of CLASP prior to beginning the general chemistry sequence was, however, significantly higher than the total course average. The grade increase for the first quarter of chemistry was 0.59 grade-points ($p<0.001$), for the second quarter 0.62 ($p<0.001$), and for the third quarter 0.43 ($p<0.03$). Although there might be a small amount of content overlap between the physics and chemistry course, we point out that these increases occurred in all three quarters of the chemistry course after only the first quarter of the CLASP course.\cite{transfer2012}

\subsubsection{MCAT scores}
In reorganizing the material for the CLASP course, we were concerned that students might not be as prepared for the MCAT so we analyzed some UC Davis students' performances on the Medical College Admissions Test (MCAT).  We use about five years of data centered on the point at which we stopped teaching Physics 5 and began teaching CLASP.  We compared our students' performance ($N = 386$ for students who took Physics 5 and $N = 347$ for students who took CLASP) on both the Physical Science and Biological portions of the MCAT.  For the Biological Science part of the test the scores ranged from 3-15 with an average of 9.71 $\pm$ 0.10 whether the students took Physics 5 or CLASP.  The Physical Science part of the test had a similar range of scores and an average of 9.26 $\pm$ 0.10 for the students who took Physics 5 and 9.42 $\pm$ 0.11 for the students who took CLASP.  This gap of 0.16 $\pm$ 0.15 suggests that the CLASP students were slightly better prepared for the MCAT but that the result is not statistically significant (for instance, a t-test has $p = 0.29$).  Nevertheless, it is important to us that the course not disadvantage our students with respect to the MCAT and certainly that seems to be the case.

\subsection{Conceptual understanding of force and motion}
For almost two decades, the Force Concept Inventory (FCI), a multiple-choice exam focusing on Newton's laws, has been a standard way for the physics education research community to measure student conceptual learning gains in introductory physics courses.\cite{Hest92}  The Force Concept Inventory (pre-test at beginning of CLASP A and post-test at the end of CLASP B) was given to four different groups of students (total of 898 students) in 1999-2001 resulting in an average pretest score of 31\% correct and an average normalized gain of 0.39 $\pm$ 0.01.  It is probably not surprising that this is well above the range associated by Hake\cite{Hake} with traditional courses and in the middle of the range of ``interactive engagement'' courses.  The FCI was never administered to the Physics 5 class, but we have no reason to assume that the class differs from other traditional lecture-based classes, which have gains of $0.24 \pm 0.03$. The normalized gain in CLASP is especially noteworthy, because only 1/3 of the second quarter (about 3.3 weeks) is devoted to motion, forces, Newton's laws, and linear momentum.

\subsection{Attitudes toward physics}
Over the past two decades, research\cite{RedishMPEX, AdamsCLASS} has shown that a majority of students leave introductory physics classrooms not only confused about the conceptual content of physics, but also about the nature of scientific knowledge.  These ideas are epistemological in nature and the implicit epistemological message sent in many traditional classrooms is apparently not what we want\cite{Schommer} our students to learn.  Indirectly, the students appear to be encouraged toward approaches to learning such as rote memorization and dissuaded from reconciling their everyday experiences with the content presented in the course to form a coherent worldview.

Several attitudinal surveys that provide information on student epistemologies have been developed\cite{RedishMPEX, AdamsCLASS} to categorize these beliefs.  In general, these surveys consist of a set of statements, such as ``Knowledge in physics consists of many pieces of information, each of which applies primarily to a specific situation.'', with which the student is asked to agree or disagree.  Student responses are coded as favorable (matching what an expert would say), as unfavorable, or as neutral.  The surveys are given twice, once at the beginning of instruction and once at the end and movement in student responses are classified as towards or away from expert-like beliefs (gains or losses).

Results from these sorts of surveys have been collected from many large lecture classes, across a variety of educational institutions, and from both `reformed' and traditional introductory physics classes.  Results are fairly consistent.  In general, unless the class has focused explicitly on addressing epistemologies,\cite{RedishHammerMPEXII} after one semester of physics instruction, student populations tend to move \emph{away} from the experts in their opinions (even for most `reformed' classes).\cite{RedishMPEX, AdamsCLASS}

In the Fall quarter of 2008 we administered the MPEX-II (Maryland Physics Expectations Survey)\cite{RedishHammerMPEXII} to two separate CLASP courses (about 600 students) of CLASP A.  The results showed that the student epistemologies in this set of classes were statistically unchanged over the course of the quarter: favorable fraction of responses changed from 0.46 $\pm$ 0.01 to 0.47 $\pm$ 0.01 and the unfavorable fraction changed from 0.27 $\pm$  0.01 to 0.28 $\pm$ 0.01.  Thus, unlike most standard classes and even many reformed Physics classes whose students seem to end the course with less expert epistemologies, this CLASP class seems to leave the students epistemological ideas unchanged on average.  The MPEX-II did not exist at the time Physics 5 did, so comparisons of CLASP are by necessity to the national sample, which, as previously mentioned, shows a decline in expert-like opinions, even for most reformed classes.

\section{Summary and conclusions}
Although the CLASP course represents a radical departure from traditional instruction, the series of introductory physics courses discussed in this paper have been fully institutionalized at UC Davis. The course has outlasted the people who originally developed it, and it continues to be positively received by faculty, and strongly supported by campus administrators. In one sense, the CLASP course serves as an existence proof that the kind of pedagogical \emph{setting} that is implicated from several decades of research from education, physics education, and cognitive science can actually be implemented in a large research university setting. As mentioned in the introduction, it basically provides a studio-physics like experience for nearly 2000 students per year and does so using no more instructional resources than a traditionally taught large lecture course with multiple lab and discussion sections. Also, even though our exams ask students to discuss their ideas in writing, time spent by TAs grading the quizzes and exams is roughly the same as spent by TAs in grading standard problems for homework and exams in the course for engineering majors. The CLASP course, including all of its pedagogical approaches, has been successfully implemented in a much smaller setting at California State University, San Marcos, where one faculty instructor teaches one section with mini-lectures interspersed in the DL setting as required.\cite{SanMarcos}

It is not only the DL setting that is important, of course. There is evidence from a variety of sources that the combination of pedagogical approaches incorporated in CLASP lead to an improvement in student performance in subsequent courses. Our strong suspicion, however, it that it is the attempt to have \emph{all} aspects of the course support the goal of sense making that is critical. Our students leave the series of CLASP courses better prepared for their later studies (MCAT and GPA comparison of Physics 5 to CLASP), with acceptable physics knowledge (FCI comparison of CLASP to national sample), and with more expert epistemologies than they would from a standard class (MPEX-II comparison to national sample). 

The CLASP course provides opportunities for training new faculty and new graduate TAs consistent with the best practices in interactive-engagement courses, precisely those best practices that earned CLASP the distinction of being designated an `exemplary practice' course in the AP redesign study (see section \ref{reviews}).  It provides a viable vehicle for faculty at a research university under very real time constraints to try out active-learning teaching approaches.

The CLASP series of courses is a work in progress, but more recently, the majority of work has shifted from fine-tuning the student activities to improve learning how to adjust the amounts of time spent on the various physics models. We have always been focused on the types of physics that our bioscience students will need to understand and use in their later courses and careers. We are guided in our efforts to make the CLASP course more relevant for our students by ongoing conversations with biological science faculty at UC Davis as well as the recent reports on undergraduate education for bio-science\cite{BIO2010} and premed\cite{AAMCHHMI} students and more recently by the ongoing ``Reforming the Introductory Physics Course for Life Science Majors'' sessions at AAPT meetings. Because our DL student activities, rather than a textbook, drive the course we can readily change the course content and rapidly adjust the weights given to the various models of the course in response to clearer understandings of what is most appropriate for these students.

Finally, as the activity development efforts wind down, our Physics Education Group is freed to use the resulting curriculum and pedagogical approaches as a learning laboratory to investigate fundamental issues of teaching and learning. In particular, we want to better understand just what it is about the combination of pedagogical approaches used in CLASP that lead to the improved performance we observe in students subsequent courses.

At this time, the CLASP series is published by Hayden McNeil as ``College Physics: A Models Approach.''  Anyone interested in learning more about or obtaining some CLASP materials may contact David Webb (webb@physics.ucdavis.edu) or Wendell Potter (potter@physics.ucdavis.edu).

\begin{acknowledgments}
The authors thank the National Science Foundation for providing the intial funding for this work under Grant No. DUE-9354528.  They also thank the many faculty, postdoctoral students, temporary lecturers and graduate students who have helped improve the CLASP courses over the years of its existence. They also acknowledge the initial and continuing support of the UC Davis Physics Department and Administration for this effort.

\end{acknowledgments}


\begin{thebibliography}{5}
\bibitem{Footnote}By this we mean, for instance, that the students are operating at the higher levels of Bloom's Taxonomy.\cite{Bloom}
\bibitem{footnoteA}The students taking this class are predominantly of sophomore or junior standing and have usually already completed introductory courses in chemistry and bilogy: they will also have completed a one-year terminal calculus course.
\bibitem{Footnote01}As an example of this kind of organization see the ``Six Ideas that Shaped Physics'' series of texts.
\bibitem{Wheeler}The spirit behind WHEELER’S FIRST MORAL PRINCIPLE – ``Never make a calculation until you know the answer.'' (from Chap.1 of Spacetime Physics by Taylor and Wheeler) is somewhat appropriate here since we insist that the students do a lot of thinking, drawing, and/or diagramming before any algebraic calculation.
\bibitem{Footnote02}Later in this paper we describe more about our instructor training and professional development.
\bibitem{Footnote03}We may even motivate the students to develop these skills by reminding them that they are practicing the argument skills that they will use on exams.
\bibitem{Footnote04}In all of the following quantitative measures, the error estimates that we quote are the standard error of the mean.
\bibitem{Footnote05}It is difficult today to determine exactly which sections of classes had ``interactive learning'' lab experiments and which did not and also whether those labs continued when the CLASP developers stopped their own experiments in the Physics 5 seris and began the CLASP courses.
\bibitem{Footnote50}For our purposes here ``bioscience'' majors are those for which the introductory physics series is required by the major, ``non-bioscience'' majors are those for which neither the Physics 5 nor Physics 7 is required.  About 2\% of the graduates could not easily be placed into one of these two categories generally because they graduated with a major that allowed but did not require our standard intro-physics series for bioscientists.
\bibitem{Footnote05A}It was not always possible to decide who was admitted as a Freshman so we have used a cutoff of 170 UCD quarter units or more (at graduation) as a proxy for ``admitted as a Freshman''.  In the years for which we can compare the proxy definition with the actual status, we find that fewer than 1\% of the students in this group are transfer students.
\bibitem{Footnote05B}It was not always possible to decide who was admitted as a transfer student so we have used a cutoff of 115 UCD quarter units or fewer (at graduation) as a proxy for ``admitted as a Junior transfer student''.  In the years for which we can compare the proxy definition with the actual status, we find that fewer than 1\% of the students in this group were admitted as a Freshman.
\end{thebibliography}
\end{document}